\documentclass[sigconf,review=false,authorversion=false,nonacm=true]{acmart}

\setcopyright{none}
\usepackage{comment}
\usepackage[utf8]{inputenc}
\usepackage{hyperref}
\usepackage{algorithm}
\usepackage[noend]{algpseudocode}
\usepackage{graphicx}
\usepackage{textcomp}
\usepackage{multicol}
\usepackage{tikz}
\usetikzlibrary{backgrounds}
\usetikzlibrary{arrows, positioning, shapes, arrows.meta}
\usetikzlibrary{patterns}
\usepackage{xcolor}
\usepackage{xspace}
\usepackage{mathpartir}
\usepackage{colortbl}
\usepackage{mdframed}
\usepackage{balance}
\usepackage{tabularx}
\usepackage{subcaption}
\usepackage{caption}
\usepackage{svg}
\usepackage{enumitem}
\usepackage{wrapfig}
\usepackage{pgfplots}
\usepgfplotslibrary{statistics}
\pgfplotsset{compat=1.18}
\usepackage[noabbrev]{cleveref}
\usepackage{multirow}
\usepackage{booktabs}
\usepackage[flushleft]{threeparttable}
\usepackage[most]{tcolorbox}
\usepackage{rotating}

\usepackage{listing}
\usepackage[normalem]{ulem}
\useunder{\uline}{\ul}{}
\usepackage{balance}

\usepackage{colortbl}
\usepackage{array}

\newcolumntype{H}{>{\columncolor{lightgray}}r}

\newcommand\constr{$\texttt{c}$\xspace}
\newcommand\nc{$\texttt{nc}$\xspace}

\newcommand\cspaceeq{$\texttt{c-space}^=$\xspace}
\newcommand\cspaceall{$\texttt{c-space}^*$\xspace}
\newcommand\ccomment{$\texttt{c-comment}$\xspace}
\newcommand\cdotted{$\texttt{c-dotted}$\xspace}

\newcommand{\LM}{L_{\mathcal{M}}}
\newcommand{\LT}{L_T}
\newcommand{\LC}{L_C}

\makeatletter
\newcommand\footnoteref[1]{\protected@xdef\@thefnmark{\ref{#1}}\@footnotemark}
\makeatother

\newcommand{\includediagram}[2][]{%
	\IfFileExists{#2}
	{}
	{
		\immediate\write18{make #2}
	}
	\includegraphics[#1]{#2}
}

\newcommand{\includepuml}[1][]{%
	\IfFileExists{#1}
	{}
	{
		\immediate\write18{make #1}
	}
	\input{#1}
}

\definecolor{lightgray}{rgb}{.9,.9,.9}
\definecolor{darkgray}{rgb}{.4,.4,.4}

\definecolor{ghBackground}{HTML}{FFFFFF}
\definecolor{ghText}{HTML}{24292E}
\definecolor{ghKeyword}{HTML}{CF222E}
\definecolor{ghType}{HTML}{0550AE}
\definecolor{ghString}{HTML}{0A3069}
\definecolor{ghNumber}{HTML}{116329}
\definecolor{ghComment}{HTML}{6E7781}
\definecolor{ghFunc}{HTML}{8250DF}

\lstdefinestyle{githublight-ts}{
  backgroundcolor=\color{ghBackground},
  basicstyle=\ttfamily\footnotesize\color{ghText},
  commentstyle=\color{ghComment}\itshape,
  keywordstyle=\color{ghKeyword}\bfseries,
  stringstyle=\color{ghString},
  numberstyle=\color{ghNumber},
  identifierstyle=\color{ghText},
  emph={string,number,boolean,any,void,never,unknown,Promise,Record,Readonly,Partial,Pick,Omit},
  emphstyle=\color{ghType},
  emph={[2]function,return,const,let,var,if,else,for,while,import,from,export,extends,implements,interface,class,new,public,private,protected,async,await,try,catch,finally,throw},
  emphstyle={[2]\color{ghKeyword}\bfseries},
  emph={[3]console,Math,Date,Array,Object,String,Number,Map,Set},
  emphstyle={[3]\color{ghFunc}},
  showstringspaces=false,
  breaklines=true,
  tabsize=2,
  rulecolor=\color{gray!20},
  frameround=tttt,
  numbers=left,
  numbersep=3pt,
  numberstyle=\tiny\color{black}
}

\lstdefinelanguage{JavaScript}{
	keywords={typeof, new, true, false, catch, function, return, null, catch, switch, var, if, in, while, do, else, case, break},
	keywordstyle=\color{blue}\bfseries,
	ndkeywords={class, export, boolean, throw, implements, import, this},
	ndkeywordstyle=\color{darkgray}\bfseries,
	identifierstyle=\color{black},
	sensitive=false,
	comment=[l]{//},
	morecomment=[s]{/*}{*/},
	commentstyle=\color{purple}\ttfamily,
	stringstyle=\color{red}\ttfamily,
	morestring=[b]',
	morestring=[b]"
}

\definecolor{tomlcomment}{HTML}{6A737D}
\definecolor{tomlstring}{HTML}{032F62}
\definecolor{tomlkey}{HTML}{005CC5}
\definecolor{tomlsection}{HTML}{6F42C1}
\definecolor{tomlnumber}{HTML}{B31D28}
\lstdefinelanguage{TOML}{
  sensitive=true,
  morecomment=[l]{\#},
  morestring=[b]",
  morestring=[b]',
  alsoletter={-},
}
\lstdefinestyle{tomlstyle}{
  language=TOML,
  basicstyle=\fontsize{7.2}{7.7}\fontfamily{fvm}\selectfont,
  commentstyle=\itshape\color{tomlcomment},
  stringstyle=\color{tomlstring},
  numbers=left,
  showstringspaces=false,
  breaklines=true,
  breakatwhitespace=true,
  columns=fullflexible,
  keepspaces=true,
  frame=none,
  upquote=true,
}

\begin{document}

\title{The Alignment Problem in Constrained Code Generation}

\author{Matteo Biagiola}
\email{matteo.biagiola@{usi,unisg}.ch} %
\orcid{0000-0002-7825-3409}
\affiliation{
	\institution{University of St.\texorpdfstring{\,}{ }Gallen and Università della Svizzera italiana (USI)}
	\streetaddress{Torstrasse 25}
	\city{St.\texorpdfstring{\,}{ }Gallen and Lugano}
	\state{SG and TI}
	\postcode{9000}
	\country{Switzerland}
}

\author{Jahrim Gabriele Cesario}
\email{jahrimgabriele.cesario@unisg.ch}
\orcid{0009-0000-1448-5676}
\affiliation{
	\institution{University of St.\texorpdfstring{\,}{ }Gallen}
	\streetaddress{Torstrasse 25}
	\city{St.\texorpdfstring{\,}{ }Gallen}
	\state{SG}
	\postcode{9000}
	\country{Switzerland}
}

\author{Luca Di Grazia}
\email{work@lucadigrazia.com}
\orcid{0000-0002-5306-8645}
\affiliation{
	\institution{University of St.\texorpdfstring{\,}{ }Gallen}
	\streetaddress{Torstrasse 25}
	\city{St.\texorpdfstring{\,}{ }Gallen}
	\state{SG}
	\postcode{9000}
	\country{Switzerland}
}

\author{George Zakhour}
\email{george.zakhour@unisg.ch}
\orcid{0009-0000-5042-1207}
\affiliation{
	\institution{University of St.\texorpdfstring{\,}{ }Gallen}
	\streetaddress{Torstrasse 25}
	\city{St.\texorpdfstring{\,}{ }Gallen}
	\state{SG}
	\postcode{9000}
	\country{Switzerland}
}

\author{Guido Salvaneschi}
\email{guido.salvaneschi@unisg.ch}
\orcid{0000-0002-9324-8894}
\affiliation{
	\institution{University of St.\texorpdfstring{\,}{ }Gallen}
	\streetaddress{Torstrasse 25}
	\city{St.\texorpdfstring{\,}{ }Gallen}
	\state{SG}
	\postcode{9000}
	\country{Switzerland}
}

\begin{abstract}
	Large Language Models (LLMs) have demonstrated strong capabilities in code generation, but their outputs frequently contain syntax or type errors that result in compilation failures. 
Constrained decoding has been proposed as a solution to mitigate compilation errors by construction, improving functional correctness as a byproduct.
However, previous works overlook a critical aspect of constrained decoding: the \textit{alignment} between constrainer (e.g., types), language model and the target specification language (e.g., TypeScript).
Misalignment is caused by the constrainer being incomplete--rejecting programs that belong to the target--or unsound--allowing programs that are not part of the target.
The \textit{bias} created by incompleteness distorts the language model distribution, and can be detrimental for code generation.

We evaluate this hypothesis using seven language models, two target languages, two constrainers, enforcing types and syntax during decoding, and we study how language models react to varying levels of incompleteness.
On three benchmarks, when the constrainer is incomplete, unconstrained decoding significantly outperforms constrained decoding in terms of functional correctness.
Incompleteness pushes the model into low-probability regions of the program space, causing the generation to frequently time out, and reducing functional correctness by up to 97\%.

These contributions make the community aware of the negative effects of misalignment in constrained decoding, and provide quantitative insights on how to design constrainers that are beneficial for code generation systems with formal guarantees.

\end{abstract}

\maketitle

\section{Introduction}
\label{sec:introduction}

Large Language Models (LLMs) showed significant capabilities in generating code~\cite{chen2021codex}. 
They excel at synthesizing programs from natural language, translating between languages, and even repairing buggy code~\cite{hou2024large}. 
Despite these advances, LLM-generated code often contains errors~\cite{ferreira2025}: generated programs frequently fail to compile due to syntax mistakes, type violations, or other logical flaws. 
In statically-typed languages like TypeScript, type-check failures dominate these errors~\cite{rastogi2015, chow2024pyty}. 
Indeed, only about 6\% of compilation errors in generated TypeScript code are purely syntactic, while roughly 94\% arise from violating the type system~\cite{muendler2025typeconstrained}. 
This observation suggests that incorporating type-system constraints could substantially improve output correctness~\cite{molinellillms}.

A promising technique to reduce such errors is constrained decoding, which enforces a language's formal rules during generation, providing correctness guarantees unlike prompting~\cite{muendler2025typeconstrained}. 
Constrained decoding~\cite{agrawal2023mgd, geng-etal-2023-grammar, muendler2025typeconstrained, park2024grammaraligned, willard2023efficientguidedgenerationlarge} refers to the class of decoding techniques for generative language models in which the token selection process is restricted to outputs that satisfy predefined formal constraints~\cite{ugare2024syncode}. 
These constraints may derive from syntactic grammars, type systems, symbolic rules, or semantic validators, and are enforced incrementally during generation~\cite{nagy2026chopchop}. 
By pruning inconsistent tokens and ensuring that only constraint-compliant continuations are explored, constrained decoding guarantees structural validity (e.g., well-formed code or data formats) and can improve reliability over post-generation validation~\cite{molinelli2025, karimipour2025llm}.

\paragraph{Significance} 
To investigate the effectiveness of constrained decoding we first conduct a preliminary study (RQ0) that compares this technique against unconstrained decoding.
To replicate and extend previous results in this space, we consider type-constrained decoding in TypeScript~\cite{muendler2025typeconstrained}, comparing against unconstrained decoding (i.e., type checking the fully generated program). 
Surprisingly, our results show that unconstrained decoding significantly outperforms type-constrained decoding in \emph{functional correctness}. 
We considered two popular code generation benchmarks, across six LLMs, three temperature settings, and multiple repetitions with statistical tests.
This result questions the effectiveness of constrained decoding and demands a robust characterization of the conditions in which constrained decoding is really effective. 

\paragraph{Methodology.} 
To explain these failures, we introduce the concept of \textit{alignment}.
Alignment captures the mutual positioning among the three core entities involved in the code generation task: 
(1)~the model language $L_{\mathcal{M}}$, i.e., the set of programs  generated by the language model $\mathcal{M}$; 
(2)~the target language $L_T$, i.e., the set of desired programs;
and 
(3)~the constrained language $L_C$, i.e., the set of programs permitted by the user-defined constraints during decoding.

The central thesis of this paper is that constrained decoding is beneficial to code generation \emph{only} when there is alignment between these three languages. 
Moreover, \emph{misalignment} explains the cases where constrained decoding is ineffective. 
This conceptualization is highly actionable: researchers and practitioners can estimate when constrained decoding will be ineffective, and are guided by concrete ways to either improve $L_C$ to align it with $L_T$, or to mitigate misalignment by finetuning $L_{\mathcal{M}}$ to align it with $L_C$.

To systematically evaluate our hypothesis, we design three research questions that study the root cause of the problem and evaluate the mitigation strategies we propose to restore alignment and make constrained decoding effective:

\begin{itemize}
    \item \textbf{RQ1 (The Cost of Misalignment).} \textit{How does misalignment between the constrainer and the target language negatively impact code generation?} 
    This question shows the problem, evaluating our hypothesis by quantifying the "incompleteness bias" to explain why current constrained decoding methods underperform.
    
    \item \textbf{RQ2 (Restoring Alignment).} \textit{To what extent can functional correctness of constrained decoding be recovered by aligning the LLM's distribution to the constrainer?} 
    This tests our first proposed mitigation strategy, evaluating whether finetuning the model to comply with constraints mitigates the detrimental incompleteness bias.
    
    \item \textbf{RQ3 (Complete Alignment).} \textit{How does constrained decoding perform under complete alignment?} 
    This question evaluates the true potential of constrained decoding in a scenario where the constrainer is complete.
\end{itemize}

\begin{figure}[t]
  \centering
  \resizebox{\columnwidth}{!}{%
  \begin{tikzpicture}[>=Stealth, every node/.style={font=\small\ttfamily}]
    \tikzstyle{valid}=[draw=green!50!black, rectangle, rounded corners=2pt, fill=green!10, minimum height=7mm, align=center, thick]
    \tikzstyle{invalid}=[draw=red!60!black, rectangle, rounded corners=2pt, fill=red!15, minimum height=7mm, align=center, thick]
    \tikzstyle{label}=[font=\scriptsize\sffamily]

    \node[valid] (func) {function};
    \node[valid, right=1.5cm of func] (main) {main() \{};

    \node[invalid, right=2cm of main, yshift=1.2cm] (helper) {helper();};
    \node[valid, right=2cm of main, yshift=-1.2cm] (const) {const x =};

    \draw[->, thick] (func) -- (main) node[midway, above, label] {$P=0.98$};
    
    \draw[->, thick] (main) -- (helper) node[midway, above left, label] {$P=0.90$};
    \draw[->, thick] (main) -- (const) node[midway, below left, label] {$P=0.10$};

    \node[draw=orange!80!black, fill=orange!15, rectangle, right=1.2cm of const, minimum height=8mm, minimum width=8mm, rounded corners=2pt, thick] (unif) {\texttt{\textbackslash n\textbackslash n\textbackslash n...}};
    
    \draw[->, thick, dashed] (const) -- (unif) node[midway, above, label] {forced path};

    \node[right=0.15cm of helper, label, text=red!80!black, align=left] {\textbf{Rejected by $C$}\\(Forward ref unsupported)};
    \node[above=0.1cm of unif, label, text=orange!90!black, align=center] {Confidence\\Collapses};

  \end{tikzpicture}
  }
  \caption{\emph{Incompleteness bias} in TypeScript. The model favors a valid forward reference after generating a function signature, but the \emph{incomplete} constrainer ($C$) rejects it, leading to a low-probability path and text degeneration.}
  \label{fig:ts-bias}

\end{figure}
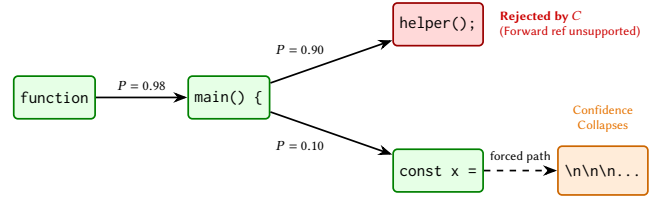

\paragraph{Results.} 
We identify that the detrimental effect of constraining observed in RQ0 stems from an \emph{incompleteness bias} (demonstrated in \autoref{fig:ts-bias}): a mismatch where an incomplete constrainer ($C$) forces the LLM into unlikely program spaces, causing timeouts and out-of-tokens events.
Aligning the language model to the constrainer through finetuning mitigates the incompleteness bias, reducing timeouts and out-of-tokens events by 53---73\% and 74---94\% respectively.
As a result, the gap between the functional correctness of constrained and unconstrained decoding solutions narrows, with the two being equivalent in some cases.
Through controlled experiments on syntax constraining with varying levels of constrainer completeness, we find that language models are highly sensitive to this factor. 
When the constrainer is complete, constrained decoding significantly outperforms its unconstrained counterpart (by 2.5\% for a 32B model and up to 54\% for a 2B model). 
When incomplete, performance drops, by up to 97\% relative to the complete-constraining-model baseline.
These results confirm the role of alignment in determining the effectiveness of 
constrained decoding.

\paragraph{Contributions} In summary, the contributions of the paper are:
\begin{itemize}
    \item A large-scale empirical study (\textbf{RQ0}) across six LLMs and two code generation benchmarks that shows that current constrained decoding approaches are consistently outperformed by unconstrained decoding in functional correctness, also on small LLMs.
    
    \item The concept of \textit{alignment} among the model language, the target language, and the constrained language. 
    This framework explains how misalignment due to incompleteness pushes the LLM into uncharted territory, generating unlikely solutions and causing an \textit{incompleteness bias} (\textbf{RQ1}).
    
    \item Empirical evidence that finetuning the LLM on constraint-model data harmonizes its distribution and effectively mitigates the incompleteness bias (\textbf{RQ2}), reducing timeouts and out-of-tokens events by 53--73\% and 74--94\%, respectively. 
    Moreover, we show the effectiveness of a complete constrainer (boosting performance by up to 54\%), whereas incomplete constraints can degrade functional correctness by up to 97\% (\textbf{RQ3}).
    
\end{itemize}

These contributions make the community aware of the negative effects of misalignment in constrained decoding, and give quantitative insights on how to design constrainers that are beneficial for code generation systems with formal guarantees.

\section{Background} \label{sec:background}

Building on prior work from \citet{nagy2026chopchop} and \citet{muendler2025typeconstrained}, we introduce the formal framework of the decoding process of language models (LLMs) for token-by-token code generation and its extensions to enforce syntactic and type-level constraints.

\subsection{Token Vocabulary and Language Models} 

Let $\Sigma$ denote a finite \emph{alphabet} of characters and $\Sigma^*$ is the set of all strings over $\Sigma$. Then, a \emph{language} $L \subseteq \Sigma^*$ is a subset of strings that are valid programs by some syntactic and semantic rules.

Since LLMs produce \emph{tokens} rather than raw characters, we introduce a token \emph{vocabulary} $\mathcal{V}$, where each token $t \in \mathcal{V}$ is a finite sequence of symbols drawn from $\Sigma$.
We assume every string $\omega \in \Sigma^*$ can be represented as a finite sequence of tokens $t_1, \ldots, t_n \in \mathcal{V}^*$. 
Formally, a \emph{language model} is a conditional probability function
$\mathcal{M} : \mathcal{V}^* \rightarrow \Delta(\mathcal{V})$,
that, given a prefix $t_1, \ldots, t_{i-1} \in \mathcal{V}^*$, produces a probability distribution over the next token $t_i \in \mathcal{V}$.
Using a language model, we generate a sequence by iteratively sampling each token $t_i$ from the model's distribution conditioned on the preceding context $t_1, \ldots, t_{i-1}$. 
This procedure, known as \emph{decoding}, terminates when the model emits the special end-of-sequence token \texttt{EOS}.

\subsection{Unconstrained Decoding}

Unconstrained decoding (\nc) represents the standard autoregressive generation  process of Large Language Models~\cite{vaswani2017attention} and a  fundamental baseline for code generation. 
Given a language model $\mathcal{M}$, an initial prompt $p_0$, a temperature setting $\tau$, and a generation budget $B$ (e.g., time and token limit), the model iteratively samples the next token based on its predicted probability distribution. 
At each step, the sampled token is appended to the prompt for the next iteration. 
When \texttt{EOS} is sampled or the budget $B$ is exhausted, the loop terminates and the final program is extracted for evaluation. 

The temperature parameter $\tau$ controls the randomness associated with model sampling. 
A temperature of 0 corresponds to greedy sampling, where the token with the highest probability is sampled from the model at each step. 
Higher temperatures correspond to higher probabilities of sampling less likely tokens.

\subsection{Constrained Decoding}
Let $\varphi$ denote a constraint over programs, such as syntactic well-formedness or type correctness. 
Constrained decoding (\constr) restricts generation by filtering out infeasible token continuations that violate $\varphi$. 
Early approaches focus on syntactic validity~\cite{dong2024xgrammar,ugare2025syncode}, while 
\citet{muendler2025typeconstrained} also model static typing rules, noting that most compilation failures stem from typing rather than parsing errors.

\begin{algorithm}[t]
\caption{Constrained Decoding Algorithm~\cite{muendler2025typeconstrained}}
\label{alg:constrained-decoding}
	\footnotesize
\begin{algorithmic}[1]
\State \textbf{Input:} Language model $\mathcal{M}$, temperature $\tau$, prompt $p_0$, budget $B$, incremental parser $\mathcal{P_I}$, incremental type-checker $\mathcal{T_I}$
\State \textbf{Output:} Candidate program $e$
\State
\Procedure{ConstrainingDecoding}{$\mathcal{M}$, $p_0$, $\tau$, $B$, $\mathcal{P_I}$, $\mathcal{T_I}$}
\State $p \gets p_0$
\While{$B$} \Comment{Budget in terms of time and number of tokens}
	\State $token_{list} \sim \mathcal{M}(\cdot \mid p, \tau)$ \Comment{List of tokens sampled by their probability}
	\For{$t$ in $token_{list}$}
		\State $p' \gets p \cup t$
		\State $e \gets ExtractProgram(p')$ 
		\If{$\mathcal{P_I}(e)$ \textbf{and} $\mathcal{T_I}(e)$}
			\State $p \gets p'$
			\If{$t$ == \texttt{EOS}}
				\State \Return $e$
			\Else
				\State \textbf{break} \Comment{Token is valid and the generation continues}
			\EndIf
		\Else
			\State \textbf{continue} \Comment{Token is invalid and the next token is considered}
		\EndIf
	\EndFor
\EndWhile

\State $e \gets ExtractProgram(p)$ \Comment{Extract the program from the partial solution}
\State \Return $e$
\EndProcedure
\end{algorithmic}
\end{algorithm}

\autoref{alg:constrained-decoding} illustrates this type-aware approach, integrating a partial parser ($\mathcal{P_I}$) and type-checker ($\mathcal{T_I}$) into the decoding loop. 
At each generation step, a candidate token $t$ is sampled and appended to the prefix $p$. If the resulting program prefix can be completed into a well-typed program, then the token is retained.  
Otherwise, the token is rejected and a new one is sampled. Note that unconstrained decoding is recovered as a special case when the partial parser and type-checker always return true, i.e., if no constraints are enforced.

\section{Empirical Study} \label{sec:preliminary}

To compare constrained decoding with the unconstrained baseline, we focus on type-constrained decoding in TypeScript~\cite{muendler2025typeconstrained}, as it is the most advanced constrainer available in the literature.
We conduct an empirical study guided by the following research question: \smallskip

\noindent
\textbf{RQ0 (Effectiveness).} \textit{For TypeScript code generation, how do type-constrained and unconstrained strategies compare in terms of functional correctness, and type safety?}

\subsection{Experimental Setup}

\subsubsection{Benchmarks}

We consider two datasets, \textsc{HumanEval}~\cite{chen2021codex} and \textsc{MBPP}~\cite{austin2021programsynthesis}, commonly used to evaluate the capabilities of LLMs on code generation.
In particular, we adopt the TypeScript datasets provided by \citet{bigcode-evaluation-harness} and available on HuggingFace\footnote{\url{https://huggingface.co/datasets/nuprl/MultiPL-E}, Accessed March 2026}.
\textsc{HumanEval} and \textsc{MBPP} consist of 159 and 390 prompts in natural language respectively.
Each prompt includes few-shot input-output examples, the function signature the model should start with, and a set of test cases to evaluate functional correctness of the generated solution.
The two datasets have been used to evaluate constrained-decoding strategies in previous work~\cite{muendler2025typeconstrained,nagy2026chopchop}, as they consist of standalone algorithmic functions. 
In fact, the existing constraining strategies only support a subset of the language (e.g.,
lacking user-defined classes) and therefore cannot be applied to repository-level benchmarks, 
such as \texttt{SWE-bench} and its variants~\cite{yang2025swesmith}.

\subsubsection{Decoding Methods}

We compare (1)~\textit{unconstrained decoding (\nc)}, i.e., standard sampling without any syntactic or type constraints, and (2)~\textit{constrained decoding (\constr)}. 
We instantiate constrained decoding in the context of TypeScript code generation, considering the approach by \citet{muendler2025typeconstrained} that guides the generation using syntax and types. 
The implementation supports the majority of features of the TypeScript language, excluding forward references, user-defined types, imports, and type inference.

We fix a generation budget $B$ of either 300 seconds or 1000 tokens on a task instance. 
We return the partial program at timeout, and count it as a failure for functional 
metrics while recording its compile status for error-rate analyses.
Both compilation and test suite execution have a 5 seconds time limit.

\begin{table*}[t]
	\centering
	\caption{\textbf{RQ0.} Type-constrained (\constr) vs unconstrained decoding (\nc) on \textsc{HumanEval} and \textsc{MBPP} benchmarks. 
	Bold values indicate a statistical significance (Wilcoxon at $\alpha = 0.05$) difference, while values are underlined when the effect size ($\hat{A}_{12}$) is large.}
	\label{tab:rq0-results-humaneval-mbpp}
	\small
	\setlength{\tabcolsep}{6.4pt}
	\renewcommand{\arraystretch}{1}
	\begin{tabular}{lllrrrrrrrrrrrr}
		\toprule

		\multirow{3}{*}{\textbf{Model}} & \multirow{3}{*}{\textbf{Dataset}} & \multirow{3}{*}{\textbf{Strategy}} & \multicolumn{3}{c}{\multirow{2}{*}{\textbf{FC $\uparrow$ (\%)}}} & \multicolumn{3}{c}{\multirow{2}{*}{\textbf{TSC $\uparrow$ (\%)}}} & \multicolumn{6}{c}{\textbf{Timeout $\downarrow$ (\%)}} \\

		\cmidrule(r){10-15}

		&  &  & \multicolumn{3}{c}{} & \multicolumn{3}{c}{} & \multicolumn{3}{c}{\textbf{Tokens}} & \multicolumn{3}{c}{\textbf{Time}} \\

		\cmidrule(r){10-12}
		\cmidrule(r){13-15}

		&  &  & \multicolumn{1}{c}{\textit{$\tau$=0.1}} & \multicolumn{1}{c}{\textit{$\tau$=0.5}} & \multicolumn{1}{c}{\textit{$\tau$=1.0}} & \multicolumn{1}{c}{\textit{$\tau$=0.1}} & \multicolumn{1}{c}{\textit{$\tau$=0.5}} & \multicolumn{1}{c}{\textit{$\tau$=1.0}} & \multicolumn{1}{c}{\textit{$\tau$=0.1}} & \multicolumn{1}{c}{\textit{$\tau$=0.5}} & \multicolumn{1}{c}{\textit{$\tau$=1.0}} & \multicolumn{1}{c}{\textit{$\tau$=0.1}} & \multicolumn{1}{c}{\textit{$\tau$=0.5}} & \multicolumn{1}{c}{\textit{$\tau$=1.0}} \\

		\midrule
 
		\multirow{4}{*}{\begin{tabular}[c]{@{}l@{}}\texttt{Gemma-2} \\ \texttt{2B}\end{tabular}} & \multirow{2}{*}{\textsc{HumanEval}} & \constr & 32.0 & 31.8 & 26.6 & 90.7 & {\ul \textbf{93.2}} & {\ul \textbf{87.6}} & 3.4 & 3.1 & 2.3 & 3.0 & 3.3 & 7.2 \\
		&  & \nc & {\ul \textbf{35.7}} & {\ul \textbf{33.6}} & 28.5 & {\ul \textbf{91.8}} & 91.1 & 83.3 & 0.1 & 0.0 & 0.0 & 0.0 & 0.0 & 0.0 \\
		& \multirow{2}{*}{\textsc{MBPP}} & \constr & 44.9 & 43.1 & 38.7 & 86.0 & {\ul \textbf{88.2}} & {\ul \textbf{83.1}} & 6.7 & 6.1 & 5.0 & 2.8 & 3.4 & 7.4 \\
		&  & \nc & {\ul \textbf{50.4}} & {\ul \textbf{49.1}} & {\ul \textbf{43.3}} & {\ul \textbf{87.1}} & 85.6 & 78.7 & 0.0 & 0.0 & 0.0 & 0.0 & 0.0 & 0.0 \\ [3pt]

		\multirow{2}{*}{\begin{tabular}[c]{@{}l@{}}\texttt{Gemma-2} \\ \texttt{9B}\end{tabular}} & \multirow{2}{*}{\textsc{HumanEval}} & \constr & 50.5 & 52.2 & 53.0 & 92.7 & 94.2 & 93.1 & 3.8 & 2.8 & 0.7 & 3.8 & 2.8 & 5.0 \\
 		&  & \nc & {\ul \textbf{58.4}} & {\ul \textbf{59.2}} & {\ul \textbf{59.2}} & 92.6 & 93.1 & 94.1 & 0.0 & 0.0 & 0.0 & 0.0 & 0.0 & 0.0 \\ [3pt]

		\multirow{2}{*}{\begin{tabular}[c]{@{}l@{}}\texttt{Gemma-2} \\ \texttt{27B}\end{tabular}} & \multirow{2}{*}{\textsc{HumanEval}} & \constr & 67.6 & 67.7 & 68.4 & 94.6 & 97.3 & 96.7 & 2.4 & 0.9 & 0.3 & 2.4 & 1.4 & 2.3 \\
 		&  & \nc & {\ul \textbf{70.4}} & {\ul \textbf{71.2}} & {\ul \textbf{70.4}} & {\ul \textbf{98.3}} & {\ul \textbf{98.6}} & {\ul \textbf{97.6}} & 0.0 & 0.0 & 0.0 & 0.0 & 0.0 & 0.0 \\ [3pt]

		\multirow{4}{*}{\begin{tabular}[c]{@{}l@{}}\texttt{Qwen-2.5} \\ \texttt{32B}\end{tabular}} & \multirow{2}{*}{\textsc{HumanEval}} & \constr & 72.3 & 75.1 & 74.7 & 90.9 & 94.6 & 94.2 & 2.5 & 1.3 & 1.1 & 1.0 & 1.1 & 1.3 \\
		&  & \nc & {\ul \textbf{82.3}} & {\ul \textbf{81.9}} & {\ul \textbf{81.5}} & {\ul \textbf{98.1}} & {\ul \textbf{97.6}} & {\ul \textbf{97.2}} & 0.0 & 0.0 & 0.0 & 0.0 & 0.0 & 0.0 \\
		& \multirow{2}{*}{MBPP} & \constr & 70.1 & 73.8 & 72.2 & 87.5 & 93.5 & 92.3 & 2.5 & 1.5 & 1.2 & 2.3 & 1.7 & 1.9 \\
		&  & \nc & {\ul \textbf{79.5}} & {\ul \textbf{79.1}} & {\ul \textbf{78.5}} & {\ul \textbf{96.4}} & {\ul \textbf{95.7}} & {\ul \textbf{94.9}} & 0.0 & 0.0 & 0.0 & 0.0 & 0.0 & 0.0 \\ [3pt]

		\multirow{2}{*}{\begin{tabular}[c]{@{}l@{}}\texttt{DSCoder} \\ \texttt{33B}\end{tabular}} & \multirow{2}{*}{\textsc{HumanEval}} & \constr & 68.0 & 68.5 & 61.8 & 95.6 & 94.7 & 91.3 & 2.9 & 3.2 & 1.6 & 3.1 & 3.2 & 5.3 \\
 		&  & \nc & {\ul \textbf{74.8}} & {\ul \textbf{75.5}} & {\ul \textbf{71.2}} & {\ul \textbf{98.3}} & {\ul \textbf{98.1}} & {\ul \textbf{96.0}} & 0.0 & 0.0 & 0.0 & 0.0 & 0.0 & 0.0 \\ [3pt]

		\multirow{2}{*}{\begin{tabular}[c]{@{}l@{}}\texttt{CodeLlama} \\ \texttt{34B}\end{tabular}} & \multirow{2}{*}{\textsc{HumanEval}} & \constr & 46.5 & 45.0 & 39.4 & {\ul \textbf{94.9}} & {\ul \textbf{95.2}} & {\ul \textbf{92.0}} & 3.3 & 2.9 & 3.2 & 1.1 & 1.0 & 3.1 \\
		&  & \nc & {\ul \textbf{50.3}} & {\ul \textbf{49.9}} & 40.8 & 92.6 & 93.3 & 85.5 & 0.1 & 0.1 & 0.0 & 0.0 & 0.0 & 0.0 \\
		
		\bottomrule
	\end{tabular}
\end{table*}

\subsubsection{Procedure}

To be compatible with the setting of \citet{muendler2025typeconstrained}, we evaluated the six LLMs in their study: three sizes from the \texttt{Gemma-2} family~\cite{gemmateam2024gemmaopenmodelsbased} (2B, 9B and 27B), and three similarly sized models from other families, DeepSeek Coder 33B (\texttt{DSCoder-33B})~\cite{guo2024deepseekcoderlargelanguagemodel}, \texttt{CodeLlama-34B}~\cite{rozière2024codellamaopenfoundation}, and \texttt{Qwen-2.5-32B}~\cite{qwen2025qwen25technicalreport}. 
We considered three different temperature settings: 0.1, 0.5, and 1.0, where higher temperature  increases randomness in the generation process~\cite{ouyang2025empiricalnondeterminism}.

To ensure rigorous and statistically valid evaluation~\cite{arcuri2014hitchhikers}, since LLMs are non-deterministic at non-zero temperatures~\cite{ouyang2025empiricalnondeterminism}, we executed 10 runs per model and decoding strategy, using different seeds, on the \textsc{HumanEval} benchmark. 
Given \textsc{MBPP}'s larger problem set, we only consider the worst and best models for \textsc{HumanEval}, namely \texttt{Gemma-2-2B} and \texttt{Qwen2.5-32B}. 
In total, the evaluation consists of (159 problems $\times$ 2 strategies $\times$ 6 LLMs $\times$ 3 temperature values $\times$ 10 repetitions) $\approx 57k$ generated \textsc{HumanEval} programs, and (390 problems $\times$ 2 strategies $\times$ 2 LLMs $\times$ 3 temperature values $\times$ 10 repetitions) $\approx 47k$ generated \textsc{MBPP} programs.

To assess whether performance differences between strategies are statistically significant, we apply the non-parametric Wilcoxon signed-rank test ($\alpha = 0.05$). 
To quantify the magnitude of these differences, we compute the Vargha-Delaney ($\hat{A}_{12}$) effect size~\cite{vargha2000critique, arcuri2014hitchhikers}. 

\subsubsection{Metrics}
\label{sec:rq0-metrics}
To measure effectiveness, we consider three metrics. 
The main metric is \textit{Functional Correctness (FC)}, which captures the percentage of generated solutions that pass all unit tests.
We then measure \textit{Type and Syntax Correctness (TSC)}, which indicates the fraction of generated solutions that pass the compilation check (using the TypeScript compiler), considering both syntax and types.
The third metric is the \textit{Timeout} rate, which measures the percentage of generated solutions that exceed the generation budget $B$, either in terms of token limit (\textit{Tokens}) or time limit (\textit{Time}).

\subsection{Results}
\label{sec:rq0-effectiveness-efficiency}

\autoref{tab:rq0-results-humaneval-mbpp} shows the average results over 10 runs for all strategies, datasets, models, and temperature settings.

For functional correctness (Columns~4---6), \textbf{unconstrained decoding (\nc) consistently outperforms constrained decoding (\constr)}, except in two cases, i.e., \texttt{Gemma-2-2B} and \texttt{CodeLlama-34B} with temperature 1.0 in the \texttt{HumanEval} benchmark, where the difference in functional correctness is not significant. 
For \nc, the highest functional correctness is achieved with the lowest temperature, i.e., 0.1, and it declines or stabilizes
when the temperature increases. 
In \textsc{HumanEval}, the highest functional correctness is $82.3\%$ achieved by \texttt{Qwen-2.5-32B}, while the lowest is $35.7\%$ achieved by \texttt{Gemma-2-2B}; the ranking between the two models is kept also for \textsc{MBPP} ($50.4\%$ vs $79.5\%$ respectively).
For \constr, functional correctness decreases significantly with increasing temperatures for three models, \texttt{Gemma-2-2B}, \texttt{CodeLlama-34B} and \texttt{DSCoder-33B}, while staying consistent or slightly improving for the others.

In terms of type-syntax correctness (Columns~7---9), \constr significantly outperforms \nc for less capable models, such as \texttt{Gemma-2-2B} and \texttt{CodeLlama-34B}.
When the temperature increases, type-syntax correctness is high for \constr, while it decreases for \nc.
However, while \constr is generally better than \nc in terms of type-syntax correctness, this fails to translate into improved functional correctness.

The lower functional correctness of \constr compared to \nc can be attributed to a non-negligible number of cases in which the constrained generation exceeds the budget, either by reaching the token (Columns~10---12) or time limit (Columns~13---15).
This limitation is more pronounced for smaller models (i.e., \texttt{Gemma-2-2B} both for \textsc{HumanEval} and \textsc{MBPP}, and \texttt{Gemma-2-9B}) and at higher temperatures.
Note however that timeouts are not exclusive to \constr; indeed, \nc can also exceed the token limit, but this occurs far less frequently than with \constr.
For \texttt{Gemma-2-2B} and \texttt{CodeLlama-34B}, \nc hits the limit in only 0.1\% of the cases at $\tau=0.1$, and a similarly low rate occurs for \texttt{CodeLlama-34B} (at $\tau \in \{0.1, 0.5\}$).
Besides timeouts, \constr only generates invalid programs 0.69\% of the cases.

\begin{tcolorbox}[boxrule=0pt,frame hidden,sharp corners,enhanced,borderline north={1pt}{0pt}{black},borderline south={1pt}{0pt}{black},boxsep=2pt,left=2pt,right=2pt,top=2.5pt,bottom=2pt]
	\textbf{RQ0 (Effectiveness)}: Unconstrained decoding consistently achieves the best functional correctness across all the configurations. 
	While type-constrained decoding improves type-syntax correctness, this does not show better functional correctness. 
\end{tcolorbox}

\section{The Alignment Problem} \label{sec:methodology}

Our preliminary study shows that constraining the output of the LLM can sometimes reduce functional correctness. 
In fact, language models are trained to approximate the true conditional  distribution of tokens in code corpora and constrainers have the effect of altering that learned distribution. 
Hence, constraining can steer decoding towards low-probability regions of the program space, resulting in  programs that are valid but statistically unrepresentative of code in the training set.
In this section, we investigate this problem, and identify the assumptions needed for effective constrained decoding.

\begin{figure*}
  \centering

  \begin{minipage}[t]{0.24\textwidth}
    \centering
    \textit{(a) Incapable Model}\\$\LM \cap \LT = \emptyset$\\[6pt]
    
    \begin{tikzpicture}[scale=0.4]
      \fill[red!40!white,  even odd rule] (4,0) circle (2);
      \fill[blue!40!white, even odd rule] (2,-2) circle (2);
      \fill[green!40!white] (0,0) circle (2);

      \begin{scope}
        \clip (0,0) circle (2);
        \fill[cyan!40!white] (2,-2) circle (2);
      \end{scope}

      \begin{scope}
        \clip (0,0) circle (2);
        \fill[yellow!40!white] (4,0) circle (2);
      \end{scope}

      \begin{scope}
        \clip (2,-2) circle (2);
        \fill[magenta!40!white] (4,0) circle (2);
      \end{scope}

      \draw (4,0) circle (2);
      \draw (2,-2) circle (2);
      \draw (0,0) circle (2);

      \node at (0, 0)   {$\mathcal{M}$};
      \node at (2, -2) {$C$};
      \node at (4, 0) {$T$};
    \end{tikzpicture}
  \end{minipage}%
  \hfill
  \begin{minipage}[t]{0.24\textwidth}
    \centering
    \textit{(b) Perfect Model}\\$\LM \setminus \LT = \emptyset$\\[6pt]
    \begin{tikzpicture}[scale=0.4]
      \fill[red!40!white,  even odd rule] (0,0) circle (2);
      \fill[blue!40!white, even odd rule] (0,-2) circle (2);
      \fill[green!40!white] (0,0) circle (1);

      \begin{scope}
        \clip (0,0) circle (1);
        \fill[yellow!40!white] (0,0) circle (2);
      \end{scope}

      \begin{scope}
        \clip (0,0) circle (2);
        \fill[magenta!40!white] (0,-2) circle (2);
      \end{scope}

      \begin{scope}
        \clip (0,0) circle (1);
        \fill[gray!40!white] (0,-2) circle (2);
      \end{scope}

      \draw (0,0) circle (2);
      \draw (0,-2) circle (2);
      \draw (0,0) circle (1);

      \node at (0, 0.5)   {$\mathcal{M}$};
      \node at (0, -3) {$C$};
      \node at (0, 1.5) {$T$};
    \end{tikzpicture}
  \end{minipage}%
  \hfill
  \begin{minipage}[t]{0.24\textwidth}
    \centering
    \textit{(c) Incoherent Constrainer}\\$\LC \cap \LT = \emptyset$\\[6pt]
        \begin{tikzpicture}[scale=0.4]
      \fill[red!40!white,  even odd rule] (4,0) circle (2);
      \fill[green!40!white, even odd rule] (2,-2) circle (2);
      \fill[blue!40!white] (0,0) circle (2);

      \begin{scope}
        \clip (0,0) circle (2);
        \fill[cyan!40!white] (2,-2) circle (2);
      \end{scope}

      \begin{scope}
        \clip (0,0) circle (2);
        \fill[magenta!40!white] (4,0) circle (2);
      \end{scope}

      \begin{scope}
        \clip (2,-2) circle (2);
        \fill[yellow!40!white] (4,0) circle (2);
      \end{scope}

      \draw (4,0) circle (2);
      \draw (2,-2) circle (2);
      \draw (0,0) circle (2);

      \node at (0, 0)   {$C$};
      \node at (2, -2) {$\mathcal{M}$};
      \node at (4, 0) {$T$};
    \end{tikzpicture}
  \end{minipage}
  \hfill
  \begin{minipage}[t]{0.24\textwidth}
    \centering
    \textit{(d) Ineffective Constrainer}\\$\LM \setminus \LC = \emptyset$\\[6pt]
    \begin{tikzpicture}[scale=0.4]
      \fill[blue!40!white,  even odd rule] (0,0) circle (2);
      \fill[red!40!white, even odd rule] (0,-2) circle (2);
      \fill[green!40!white] (0,0) circle (1);

      \begin{scope}
        \clip (0,0) circle (1);
        \fill[cyan!40!white] (0,0) circle (2);
      \end{scope}

      \begin{scope}
        \clip (0,0) circle (2);
        \fill[magenta!40!white] (0,-2) circle (2);
      \end{scope}

      \begin{scope}
        \clip (0,0) circle (1);
        \fill[gray!40!white] (0,-2) circle (2);
      \end{scope}

      \draw (0,0) circle (2);
      \draw (0,-2) circle (2);
      \draw (0,0) circle (1);

      \node at (0, 0.5)   {$\mathcal{M}$};
      \node at (0, -3) {$T$};
      \node at (0, 1.5) {$C$};
    \end{tikzpicture}
  \end{minipage}
  \caption{Irrelevant combinations of the model language $\LM$, target language $\LT$, and the constrained language $\LC$. 
  }
  \label{fig:irrelevant-cases}
\end{figure*}
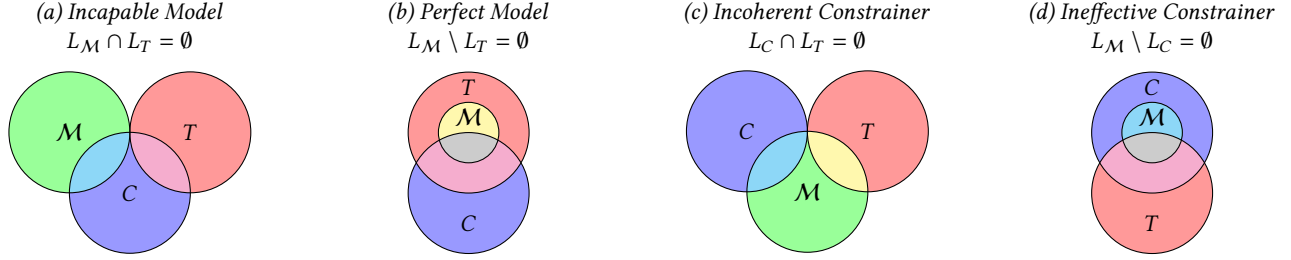

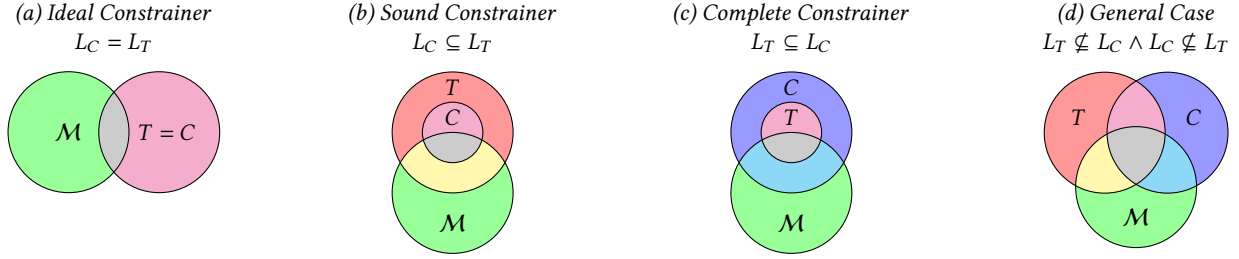
\begin{figure*}
  \centering

  \begin{minipage}[t]{0.24\textwidth}
    \centering
    \textit{(a) Ideal Constrainer}\\$\LC = \LT$\\[6pt]
    \begin{tikzpicture}[scale=0.4]
      \fill[magenta!40!white,  even odd rule] (3,0) circle (2);
      \fill[green!40!white] (0,0) circle (2);

      \begin{scope}
        \clip (0,0) circle (2);
        \fill[gray!40!white] (3,0) circle (2);
      \end{scope}

      \draw (3,0) circle (2);
      \draw (0,0) circle (2);

      \node at (0, 0)   {$\mathcal{M}$};
      \node[shift={(0.0:0.1)}] at (3, 0) {$T=C$};
    \end{tikzpicture}
  \end{minipage}%
  \hfill
  \begin{minipage}[t]{0.24\textwidth}
    \centering
    \textit{(b) Sound Constrainer}\\$\LC \subseteq \LT$\\[6pt]
    \begin{tikzpicture}[scale=0.4]
      \fill[red!40!white,  even odd rule] (0,0) circle (2);
      \fill[green!40!white, even odd rule] (0,-2) circle (2);
      \fill[blue!40!white] (0,0) circle (1);

      \begin{scope}
        \clip (0,0) circle (1);
        \fill[magenta!40!white] (0,0) circle (2);
      \end{scope}

      \begin{scope}
        \clip (0,0) circle (2);
        \fill[yellow!40!white] (0,-2) circle (2);
      \end{scope}

      \begin{scope}
        \clip (0,0) circle (1);
        \fill[gray!40!white] (0,-2) circle (2);
      \end{scope}

      \draw (0,0) circle (2);
      \draw (0,-2) circle (2);
      \draw (0,0) circle (1);

      \node at (0, -3) {$\mathcal{M}$};
      \node at (0, 0.5) {$C$};
      \node at (0, 1.5) {$T$};
    \end{tikzpicture}
  \end{minipage}%
  \hfill
  \begin{minipage}[t]{0.24\textwidth}
    \centering
    \textit{(c) Complete Constrainer}\\$\LT \subseteq \LC$\\[6pt]
    \begin{tikzpicture}[scale=0.4]
      \fill[blue!40!white,  even odd rule] (0,0) circle (2);
      \fill[green!40!white, even odd rule] (0,-2) circle (2);
      \fill[red!40!white] (0,0) circle (1);

      \begin{scope}
        \clip (0,0) circle (1);
        \fill[magenta!40!white] (0,0) circle (2);
      \end{scope}

      \begin{scope}
        \clip (0,0) circle (2);
        \fill[cyan!40!white] (0,-2) circle (2);
      \end{scope}

      \begin{scope}
        \clip (0,0) circle (1);
        \fill[gray!40!white] (0,-2) circle (2);
      \end{scope}

      \draw (0,0) circle (2);
      \draw (0,-2) circle (2);
      \draw (0,0) circle (1);

      \node at (0, -3) {$\mathcal{M}$};
      \node at (0, 0.5) {$T$};
      \node at (0, 1.5) {$C$};
    \end{tikzpicture}
  \end{minipage}
  \hfill
  \begin{minipage}[t]{0.24\textwidth}
    \centering
    \textit{(d) General Case}\\$\LT \not\subseteq \LC \land \LC \not\subseteq \LT$\\[6pt]
    \begin{tikzpicture}[scale=0.4]
      \fill[green!40!white]  (270:1.2) circle (2);
      \fill[blue!40!white]    (30:1.2)  circle (2);
      \fill[red!40!white]   (150:1.2) circle (2);

      \begin{scope}
        \clip (270:1.2) circle (2);
        \fill[cyan!40!white] (30:1.2) circle (2);
      \end{scope}
      \begin{scope}
        \clip (270:1.2) circle (2);
        \fill[yellow!40!white]  (150:1.2) circle (2);
      \end{scope}
      \begin{scope}
        \clip (30:1.2) circle (2);
        \fill[magenta!40!white] (150:1.2) circle (2);
      \end{scope}

      \begin{scope}
        \clip (270:1.2) circle (2);
        \clip (30:1.2)  circle (2);
        \fill[gray!40!white] (150:1.2) circle (2);
      \end{scope}

      \foreach \angle/\label in {270/$\mathcal{M}$, 30/$C$, 150/$T$}
        \draw (\angle:1.2) circle (2)
              node[shift={(\angle:0.40)}] {\label};

    \end{tikzpicture}
  \end{minipage}
  \caption{Alignment between the model language $\LM$, target language $\LT$, and the constrained language $\LC$. 
  }
  \label{fig:venn}
\end{figure*}

\subsection{Model, Target, and Constrained Languages}
\label{sec:languages}

Three languages are involved in constrained decoding. %

First, the \textit{model language} $\LM$ is the set of ``likely'' programs that can be generated by the language model $\mathcal{M}$. 
More formally, $\LM$~includes all the programs that can be generated by some model $\mathcal{M}$ under some code generation prompt with a non-negligible  probability, i.e., above a certain threshold $\epsilon$. 
Such  $\epsilon$ filters out the programs that can be theoretically generated by $\mathcal{M}$, but are not representative of the model's learned distribution.

Second, the \textit{target language} $\LT$ is the set of programs we expect the model to generate. 
More formally, $\LT$~includes all programs that are valid  according to some language specification (e.g., Typescript), which  is the syntactic (e.g., grammar) and semantic rules (e.g., type system) that programs must satisfy to compile successfully.

Lastly, the \textit{constrained language} $\LC$ is the set of programs that are compliant with some user-defined constraining model, or \textit{constrainer}. 
More formally, $\LC$~includes all complete programs that are valid by some syntactic $\mathcal{P_I}$ and semantic $\mathcal{T_I}$ constraints (from \autoref{alg:constrained-decoding}), which are often approximating the target language. 

In fact, $\mathcal{P_I}$ and $\mathcal{T_I}$ must be defined over program prefixes to be compliant with the auto-regressive nature of language models. 
For programming languages with a complex type system, \textit{soundness} or \textit{completeness} of the constrainer is precluded~\cite{nagy2026chopchop}, either forcing to accept some invalid programs 
(i.e., false positives) or to reject some valid ones (i.e., false negatives).

The thesis of this paper is that constrained decoding is detrimental for code generation when  these three entities are not \textit{aligned}, and describe the \textit{alignment problem} in the following section.

\subsection{Language Misalignment}

We frame the alignment problem in terms of possible relationships among the three sets $\LM$, $\LT$, and $\LC$.

\paragraph{Assumptions} Three sets can overlap in $2^7$ ways, each characterized by the presence or absence of the $7$ possible intersections. 
Yet, most cases are irrelevant for code generation, so we can exclude them by introducing some reasonable assumptions on the sets.

The first assumption is that the language model $\mathcal{M}$ is \textit{capable}, i.e., it can generate programs in the target language (i.e., $\LM \cap \LT \neq \emptyset$). 
The second assumption is that $\mathcal{M}$ is \textit{imperfect}, i.e., it can generate programs that are not in the target language (i.e., $\LM \setminus \LT \neq \emptyset$). 
The last assumptions are that the constrainer is \textit{coherent}, i.e., it accepts some programs in the target language (i.e., $\LC \cap \LT \neq \emptyset$), and \textit{effective}, i.e., it rejects some in the language model (i.e., $\LM \setminus \LC \neq \emptyset$). 
Without any of these assumptions, constraining is never beneficial.

\autoref{fig:irrelevant-cases} shows examples violating each assumption. 
We represent the languages as circles: green the model language $\LM$, red the target language $\LT$, and blue the constraining language $\LC$ (for short, $\mathcal{M}$, $T$ and $C$ in the figures). 
Also, we denote in yellow the model's knowledge of the target language (i.e., $\LM \cap \LT$), cyan the model's knowledge of the constraints (i.e., $\LM \cap \LC$), magenta the adherence of the constraints to the target language (i.e., $\LC \cap 
\LT$), and gray their intersection, i.e., the alignment of the languages.

More formally, we define the \textit{alignment} $\mathcal{A}$ as the Jaccard similarity between the three languages: 

$$\mathcal{A} = \frac{|\LM \cap \LC \cap \LT|}{|\LM \cup \LC \cup \LT|}$$

Then, the \textit{alignment problem} is the problem of maximizing the alignment, changing $\LM$ and $\LC$ with a fixed target language $\LT$.

\paragraph{Ideal Cases} By these assumptions, we find the optimal alignment when $\mathcal{M}$ has been trained on the target language (i.e., $\LM \cap \LT \neq \emptyset$): the constrainer is perfectly aligned with the target language (i.e., $\LC = \LT$), as in \autoref{fig:venn}.a. 
This means that the constrainer is both sound, i.e., it allows only programs that are in $\LT$, and complete, i.e., it allows all the programs that are in $\LT$. 
Yet, such a constrainer often does not exist in practice (as discussed in
\autoref{sec:languages}).

Therefore, we can relax the condition of optimality by allowing the constrainer to be either sound or complete, so long as we add further restrictions on the language model. 
When the constrainer is sound (i.e., $\LC \subseteq \LT$), optimal alignment is restored if the language model cannot produce valid programs outside of the constrainer $\LC$ (i.e., $(\LM \cap \LT) \setminus \LC = \emptyset$).
When the constrainer is complete (i.e., $\LT \subseteq \LC$), optimal alignment is restored if the language model cannot produce invalid programs in $\LC$ (i.e., $(\LM \cap \LC) \setminus \LT = \emptyset$).

As example of complete constrainer, consider $\LT$ being Java 8, while $\LC$ supports Java 25 programs, but the language model was trained on Java 8 programs only.
Java 25 is backward-compatible ($\LT \subseteq \LC$), but some of its features (e.g., the vector API) are not supported by the Java 8 compiler. 
For the soundness case, consider the same model, but $\LT$ is Java 25 and 
$\LC$ supports Java 8 programs.

We conjecture that these are the ideal cases for constraining to be beneficial, as, from the perspective of the language model, the constrainer and the target language are \textit{aligned}.

\paragraph{General Cases} In practice, optimal alignment is rarely achieved, hence the languages fall in one of the three following general cases.

\autoref{fig:venn}.b shows the first general case: the constrainer is sound, but the language model can generate valid programs not supported by the constrainer (i.e., $(\LM \cap \LT) \setminus \LC \neq \emptyset$, the yellow region).
This is a common practical case, arising naturally from constrained decoding: constraints are enforced token-by-token, on partial programs, and certain features cannot be implemented efficiently at this granularity~\cite{loula2025syntactic}.
For instance, features such as forward references--i.e., using a variable, function, or class before it is defined--cannot be enforced at the token level without backtracking, as the validity of an earlier token may only become apparent much later in the sequence. 
As the number of backtracking steps grows, this process slows down generation and, in the worst case, converges to rejection sampling.
As a result, a constrainer may simply not support such features, even if they are valid in the target language and the language model has been trained on them.
Given the same prompt, the language model can therefore generate programs that use forward references, falling in the yellow region, alongside programs that do not, falling in the gray region.

\autoref{fig:venn}.c shows the second general case, where the constrainer is complete, but the language model can generate invalid programs that are accepted by the constrainer (i.e., $(\LM \cap \LC) \setminus \LT \neq \emptyset$, the cyan region). 
One example is a syntactic constrainer for a strongly-typed target language, which results in the constrainer accepting programs that are syntactically valid with potential type errors. 
In this case, the  language model can generate programs that are accepted by the constrainer, but are rejected by the target language.

\autoref{fig:venn}.d shows the last and most general case, where there are no empty intersections between the languages, that is the constrainer is ill-designed, being neither sound nor complete.

In these general cases, there is little alignment between $\LC$ and $\LT$, and the constrainer can be detrimental for the code generation task. 
In fact, unsoundness results in generating programs outside the target language, while incompleteness introduces
a \textit{bias} in the generation, forcing the language model to generate
programs with a reduced set of features supported by the target.
In fact, if $\mathcal{M}$ uses unsupported features with \textit{high probability} during unconstrained generation, an incomplete constrainer might steer the generation towards low probability regions of the program space,
causing text degeneration and decreasing functional correctness.

\subsection{Mitigating Misalignment}

From the most general case (\autoref{fig:venn}.d), the designer of a constrainer
should transition towards an ideal constrainer (\autoref{fig:venn}.a), progressively reducing the amount of bias.
The first choice is between completeness (\autoref{fig:venn}.c) and soundness
(\autoref{fig:venn}.b), given that having both is unrealistic in practice.

We argue, however, that soundness (\autoref{fig:venn}.b) should be prioritized over completeness to benefit from the guarantees of constrained decoding:  if the constrainer is sound, all the solutions generated under constrained decoding are guaranteed to be \textit{correct}, according to the user-defined constraints.
Then, the incompleteness bias of sound constraints (i.e., the yellow area in \autoref{fig:venn}.b) can be reduced in two ways: the language model can be \textit{trained} to generate programs that are compliant with the constrainer through finetuning, and/or the constrainer can be enlarged to include more of the target language.
In the next sections, we evaluate how the incompleteness bias can be mitigated through finetuning, and how sensitive are language models to varying degrees of incompleteness.

\section{Experimental Evaluation}\label{sec:eval}

In this section, we investigate how misalignment between constrainer and target language specification affects the code generation task. 
We formulate the following research questions:

\textbf{RQ1 (The Cost of Misalignment).} \textit{How does misalignment between the constrainer and the target language negatively impact code generation?} 
In this RQ, we study the reasons why the constrained decoding technique we evaluated in RQ0 underperforms unconstrained decoding. 
The alignment case we evaluated in RQ0 and analyze in RQ1 is most similar to \autoref{fig:venn}.b, where $\LT$ is the Typescript language, and $\LC$ is the constrainer implemented by \citet{muendler2025typeconstrained}.
$\LC \subseteq \LT$, because some of the features of $\LT$ are not supported, such as forward reference, import statements and user-defined types, and the vast majority of programs that are supported by $\LC$ are also supported by $\LT$; at the same time $\mathcal{M}$ is trained on all features of Typescript, hence the yellow intersection is not empty.
From the perspective of $\mathcal{M}$, $\LC$ and $\LT$ are \textit{misaligned}; in RQ1 we argue that this type of misalignment is detrimental for constrained decoding, and we provide quantitative evidence using Typescript and type-constrained decoding as a concrete case study.

\textbf{RQ2 (Restoring Alignment).} \textit{To what extent can functional correctness of constrained decoding be recovered by aligning the LLM's distribution to the constrainer?} 
RQ2 investigates \textit{finetuning} as a practical way to align the language model to the constrainer. 
In RQ2 we still consider type-constrained decoding in Typescript, i.e., the case in \autoref{fig:venn}.b; finetuning $\mathcal{M}$ allows us to transition towards the ideal case for a sound constrainer (i.e., \autoref{fig:venn}.a), where $\mathcal{M}$ cannot produce valid programs outside of the constrained language (i.e., $(\LM \cap \LT) \setminus \LC = \emptyset$).
We analyze the effect of finetuning on type-constrained decoding.

\textbf{RQ3 (Complete Alignment).} \textit{How does constrained decoding perform under complete alignment?} 
In RQ3, we systematically evaluate the impact of misalignment using syntax constraining in TOML.
While prior experiments on TypeScript suffer from constraint-induced bias due to the language's complexity, TOML provides the perfect sandbox to evaluate our hypothesis. 
It features a strict, unambiguous, and predominantly context-free grammar, while still maintaining high real-world relevance as it is a language adopted by modern build systems (e.g., Rust's \texttt{Cargo} and Python's \texttt{uv} and \texttt{poetry}). 
In this RQ, we first evaluate language models under a complete syntactic constrainer (i.e., \autoref{fig:venn}.c). 
Then, we systematically introduce incompleteness bias by artificially removing specific \textit{syntactic} and \textit{semantic} features, transitioning to \autoref{fig:venn}.d. 
By reducing completeness, we can isolate and quantify exactly how sensitive LLMs are to varying degrees of incompleteness bias.

\subsection{RQ1: The Cost of Misalignment}

\begin{table}[t]
	\centering
	\caption{\textbf{RQ1}. 
	Percentage of tasks where the average NLL of constrained decoding (\textbf{\constr}) is significantly greater ($\uparrow$), lower ($\downarrow$), or equivalent ($=$) to that of unconstrained decoding (\textbf{\nc}).
	}
	\label{tab:results-bias}
	\footnotesize
	\setlength{\tabcolsep}{3.8pt}
	\renewcommand{\arraystretch}{1.2}
	\begin{tabular}{llrrrrrr}
		\toprule

		\multirow{3}{*}{\textbf{Model}} & \multirow{3}{*}{\textbf{Dataset}} & \multicolumn{6}{c}{\textbf{\constr vs \nc}} \\

		\cmidrule(r){3-8}

		&  & \multicolumn{3}{c}{\textit{$\tau$=0.1}} & \multicolumn{3}{c}{\textit{$\tau$=1.0}} \\

		\cmidrule(r){3-5}
		\cmidrule(r){6-8}

		&  & \multicolumn{1}{c}{\textit{$\uparrow$ bias}} & \multicolumn{1}{c}{\textit{$=$ bias}} & \multicolumn{1}{c}{\textit{$\downarrow$ bias}} & \multicolumn{1}{c}{\textit{$\uparrow$ bias}} & \multicolumn{1}{c}{\textit{$=$ bias}} & \multicolumn{1}{c}{\textit{$\downarrow$ bias}} \\

		\midrule

		\multirow{2}{*}{\texttt{Gemma-2-2B}} & \textsc{HumanEval} & 25.79 & 72.96 & 1.26 & 13.21 & 86.79 & 0.00 \\
		& MBPP & 32.82 & 66.41 & 0.77 & 13.08 & 85.90 & 1.03 \\

		\texttt{Gemma-2-9B} & \textsc{HumanEval} & 25.79 & 72.96 & 1.26 & 24.53 & 75.47 & 0.00 \\

		\texttt{Gemma-2-27B} & \textsc{HumanEval} & 20.13 & 79.25 & 0.63 & 17.61 & 80.50 & 1.89 \\

		\multirow{2}{*}{\texttt{Qwen-2.5-32B}} & \textsc{HumanEval} & 18.24 & 77.36 & 4.40 & 18.24 & 81.76 & 0.00 \\
		& MBPP & 19.23 & 80.26 & 0.51 & 18.46 & 80.26 & 1.28 \\

		\texttt{DSCoder-33B} & \textsc{HumanEval} & 21.38 & 76.10 & 2.52 & 13.84 & 86.16 & 0.00 \\

		\texttt{CodeLlama-34B} & \textsc{HumanEval} & 16.35 & 80.50 & 3.14 & 5.03 & 93.08 & 1.89 \\

		\midrule

		\rowcolor{lightgray} \textit{Avg} & \textit{---} & \textit{22.47} & \textit{75.73} & \textit{1.81} & \textit{15.50} & \textit{83.74} & \textit{0.76} \\
		\bottomrule
	\end{tabular}
\end{table}

\subsubsection{Metrics}

To answer RQ1, we quantify how misalignment biases the underlying probability distribution by computing the \emph{normalized negative log-likelihood} (NLL) of generated programs:
$
\mathrm{NLL}(\mathcal{T}; \mathcal{M}) = -\frac{1}{n} \sum_{i=1}^{n} \log \mathcal{M}(\tau_i \mid \mathcal{T}_{<i})
$,
where $\mathcal{T}$ is a program of $n$ tokens and $\mathcal{M}$ is the language model. 
Comparing the average NLL between constrained and unconstrained decoding measures the extent of constraint-induced bias. 
Higher NLL values indicate that the constrainer forces $\mathcal{M}$ to select less likely tokens, exposing the internal disagreement between the model's learned distribution and the enforced formal rules. 

\subsubsection{Results}

\autoref{tab:results-bias} quantifies decoding bias by comparing the NLL of solutions generated under constrained (\constr) and unconstrained (\nc) decoding at $\tau \in \{0.1, 1.0\}$.
For each problem instance, we compared the NLL distributions of the 10 solutions generated by each strategy using the Wilcoxon test and Vargha-Delaney effect size.
An $\hat{A}_{12}$ above 0.5 indicates that constrained decoding produces more unnatural solutions ($\uparrow$ bias); below 0.5, the opposite holds ($\downarrow$ bias); otherwise, no meaningful difference is recorded ($=$ bias).

Bias is consistently more frequent under constrained decoding, especially at $\tau=0.1$, where up to one-third of instances show significantly higher NLL (e.g., 35.82\% for \texttt{Gemma-2-2B} on \textsc{MBPP}).
At $\tau=1.0$, stochastic sampling reduces this effect substantially: biased instances drop (e.g., from 35.82\% to 13.08\% for \texttt{Gemma-2-2B} on \textsc{MBPP}), and statistically equivalent cases grow beyond 80\% on average.
This reveals a temperature-dependent trade-off: lower temperatures amplify constraint-induced bias by restricting diversity, while higher temperatures mitigate it at the cost of increased solution variability.
Notably, more capable models (e.g., \texttt{Gemma-2-27B} and \texttt{Qwen2.5-32B}) show lower sensitivity to these effects.

\begin{figure}
    \centering
    \begin{tikzpicture}
        \begin{axis}[
            width=9cm,
            height=5cm,
            boxplot/box extend=0.4,
            ylabel={NLL},
            xtick={1,2,3,4,5},
            xticklabels={
                All\\ \constr, 
                All\\ \nc, 
                Timeout\\ \constr, 
                Correct\\ \constr, 
                Correct\\ \nc
            },
            xticklabel style={align=center, anchor=north, text width=2cm},
            tick label style={font=\small},
            label style={font=\small},
            ymin=-0.02,
            ymajorgrids=true,
            grid style={dashed, gray!40},
            boxplot/draw direction=y
        ]
        
        \addplot+[boxplot, draw=blue!60!black, fill=blue!25, mark=*, mark size=1.2pt, mark options={fill=blue!70!black}] 
            table[y index=0, col sep=comma]{./figures/nll_comparison_humaneval_gemma-2-2b_t=0.1.csv};

        \addplot+[boxplot, draw=green!60!black, fill=green!25, mark=*, mark size=1.2pt, mark options={fill=green!70!black}] 
            table[y index=2, col sep=comma]{./figures/nll_comparison_humaneval_gemma-2-2b_t=0.1.csv};
            
        \addplot+[boxplot, draw=blue!60!black, fill=blue!25, mark=*, mark size=1.2pt, mark options={fill=blue!70!black}] 
            table[y index=1, col sep=comma]{./figures/nll_comparison_humaneval_gemma-2-2b_t=0.1.csv};
        
        \addplot+[boxplot, draw=blue!60!black, fill=blue!25, mark=*, mark size=1.2pt, mark options={fill=blue!70!black}] 
            table[y index=0, col sep=comma]{./figures/nll_comparison_humaneval_gemma-2-2b_t=0.1_filtered.csv};
            
        \addplot+[boxplot, draw=green!60!black, fill=green!25, mark=*, mark size=1.2pt, mark options={fill=green!70!black}] 
            table[y index=1, col sep=comma]{./figures/nll_comparison_humaneval_gemma-2-2b_t=0.1_filtered.csv};
        
        \end{axis}
    \end{tikzpicture}
    \caption{\textbf{RQ1}. NLL for \texttt{Gemma-2-2B} under constrained (\constr) and unconstrained (\nc) decoding at $\tau=0.1$ on \textsc{HumanEval}.
	}
    \label{fig:box-plots-rq2}
\end{figure}
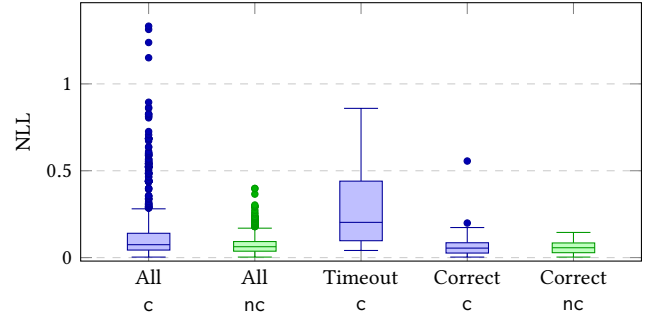

\autoref{fig:box-plots-rq2} shows NLL distributions for \texttt{Gemma-2-2B} at $\tau=0.1$ on \textsc{HumanEval} (results for \textsc{MBPP} are similar).
Overall, constrained and unconstrained distributions differ significantly, with constrained decoding producing higher NLL.
This is most pronounced for timed-out solutions, which are extremely unlikely under the model's distribution, illustrating how an incomplete constrainer can push the language model into \emph{uncharted territory}.

In contrast, when restricting to problem instances where both strategies produce a functionally correct solution, the two distributions are nearly identical: the Wilcoxon test fails to reject the null hypothesis and the effect size is negligible ($\hat{A}_{12} = 0.502$).
Similar results extend to other models.
This indicates that the bias introduced by constrained decoding in the full dataset, is largely due to an incomplete constrainer, forcing the language model to use a subset of features of the Typescript language, hence making constrained-generated solutions more unlikely.

To further validate this hypothesis, we gathered all solutions produced by the unconstrained strategy for the \texttt{Gemma-2-2B} model on \textsc{HumanEval}, across all runs and temperature settings, that successfully compiled.
Among the 4,231 unconstrained solutions that compiled, 982 failed the incremental
type-checking pass of the TypeScript constrainer.

\begin{lstlisting}[style=githublight-ts, language=JavaScript, caption={Solution generated by the unconstrained decoding for the \textsc{HumanEval} ID 107 problem. The ``!X!'' symbol at Line~5 shows where the incremental type-checker stops.}, label={lst:forwarded-reference-unconstrained}]
function even_odd_palindrome(n:number): [number,number] {
	let evenCount = 0;
	let oddCount = 0;
	for (let i = 1; i <= n; i++) {
		if (is!X!Palindrome(i) && i %
			evenCount++;
		} else if (isPalindrome(i) && i %
			oddCount++;
		}
	}
	return [evenCount, oddCount];
}
function isPalindrome(num:number): boolean {
	const str = num.toString();
	return str === str.split('').reverse().join('');
}
\end{lstlisting}

\autoref{lst:forwarded-reference-unconstrained} shows one such instance, where the incremental type checker fails to compile the solution due to an unsupported forward reference.
Under constrained decoding, the constraining mechanism blocks the call to the not-yet-defined function (``!X!'' at Line~5) and forces the model to select an existing global function instead.
In this case, the model completes the prefix \texttt{is} as \texttt{isNaN} rather than \texttt{isPalindrome}, then attempts to redefine the logic in a helper \texttt{isN} that is never called.
The result is type-safe and syntactically correct, but not functionally correct, consistent with our general findings in RQ0.
In other instances, forcing \texttt{isNaN} instead of \texttt{isPalindrome} triggers a \textit{generation loop}, where the model repeats the same token or sequence until the timeout or token limit is reached.

\begin{tcolorbox}[boxrule=0pt,frame hidden,sharp corners,enhanced,borderline north={1pt}{0pt}{black},borderline south={1pt}{0pt}{black},boxsep=2pt,left=2pt,right=2pt,top=2.5pt,bottom=2pt]
	\textbf{RQ1 (The Cost of Misalignment)}: Misalignment between the constrainer and the target language is detrimental for constrained decoding. 
	The incompleteness of the constrainer distorts the LLM's learned distribution during constrained decoding, by generating unlikely solutions and pushing the model toward \emph{uncharted territory}, reducing functional correctness.
\end{tcolorbox}

\subsection{RQ2: Restoring Alignment}

\begin{table*}[t]
	\centering
	\caption{\textbf{RQ2}. The results of \texttt{Gemma-2-2B} on \textsc{HumanEval} and \textsc{MBPP} with $\tau$=0.1, before (OR) and after (FT, gray) finetuning. Bold values indicate statistical significance between \constr and \nc (Wilcoxon at $\alpha = 0.05$) and underlined values a large effect size ($\hat{A}_{12}$).}
	\label{tab:results-finetuning}
	\small
	\setlength{\tabcolsep}{6.5pt}
	\renewcommand{\arraystretch}{0.85}
	\begin{tabular}{lllrHrHrHrH}
		\toprule

		&  &  & \multicolumn{2}{c}{} & \multicolumn{2}{c}{} & \multicolumn{4}{c}{\textbf{Timeout $\downarrow$ (\%)}} \\

		\cmidrule(r){8-11}

		&  &  & \multicolumn{2}{c}{\multirow{-2}{*}{\textbf{FC $\uparrow$ (\%)}}} & \multicolumn{2}{c}{\multirow{-2}{*}{\textbf{TSC $\uparrow$ (\%)}}} & \multicolumn{2}{c}{\textbf{Tokens}} & \multicolumn{2}{c}{\textbf{Time}} \\

		\cmidrule(r){8-9}
		\cmidrule(r){10-11}

		\multirow{-3}{*}{\textbf{Model}} & \multirow{-3}{*}{\textbf{Dataset}} & \multirow{-3}{*}{\textbf{Strategy}} & \multicolumn{1}{c}{\textit{OR ($\tau$=0.1)}} & \multicolumn{1}{c}{\textit{FT ($\tau$=0.1)}} & \multicolumn{1}{c}{\textit{OR ($\tau$=0.1)}} & \multicolumn{1}{c}{\textit{FT ($\tau$=0.1)}} & \multicolumn{1}{c}{\textit{OR ($\tau$=0.1)}} & \multicolumn{1}{c}{\textit{FT ($\tau$=0.1)}} & \multicolumn{1}{c}{\textit{OR ($\tau$=0.1)}} & \multicolumn{1}{c}{\textit{FT ($\tau$=0.1)}} \\

		\midrule

		&  & \constr & 32.0 & 32.1 & 90.7 & {\ul \textbf{99.0}} & 3.4 & 0.2 & 3.0 & 0.8 \\
		& \multirow{-2}{*}{\textsc{HumanEval}} & \nc & {\ul \textbf{35.7}} & 32.8 & {\ul \textbf{91.8}} & 97.9 & 0.1 & 0.0 & 0.0 & 0.0 \\
		&  & \constr & 44.9 & 46.6 & 86.0 & 95.1 & 6.7 & 1.7 & 2.8 & 1.3 \\
		\multirow{-4}{*}{\begin{tabular}[c]{@{}l@{}}\texttt{Gemma-2} \\ \texttt{2B}\end{tabular}} & \multirow{-2}{*}{\textsc{MBPP}} & \nc & {\ul \textbf{50.4}} & {\ul \textbf{48.5}} & {\ul \textbf{87.1}} & 95.1 & 0.0 & 0.0 & 0.0 & 0.0 \\

		&  & \constr & 72.3 & 77.6 & 90.9 & 96.2 & 2.5 & 0.0 & 1.0 & 0.4 \\
		& \multirow{-2}{*}{\textsc{HumanEval}} & \nc & {\ul \textbf{82.3}} & {\ul \textbf{82.0}} & {\ul \textbf{98.1}} & {\ul \textbf{98.6}} & 0.0 & 0.0 & 0.0 & 0.0 \\
		&  & \constr & 70.1 & 76.3 & 87.5 & 96.0 & 2.5 & 0.8 & 2.3 & 0.3 \\
		\multirow{-4}{*}{\begin{tabular}[c]{@{}l@{}}\texttt{Qwen2.5} \\ \texttt{32B}\end{tabular}} & \multirow{-2}{*}{\textsc{MBPP}} & \nc & {\ul \textbf{79.5}} & {\ul \textbf{80.1}} & {\ul \textbf{96.4}} & {\ul \textbf{98.0}} & 0.0 & 0.0 & 0.0 & 0.0 \\
		
		\bottomrule
	\end{tabular}
\end{table*}

\subsubsection{Procedure}
We finetuned the worst and best model of RQ0, namely \texttt{Gemma-2-2B} and \texttt{Qwen2.5-32B}, to reduce the bias caused by incompleteness of the constrainer, targeting the gap between programs accepted by the TypeScript compiler but rejected by the constrainer.
We focused on temperature $\tau=0.1$ which gives the highest functional correctness.

We identified hard problem instances by computing the \textit{alignment rate}, i.e. the fraction of generated solutions passing the incremental type checker, over 10 solutions per instance from RQ0.
We selected instances with an alignment rate below 10\%, resulting in 37/159 (23\%) instances
for \textsc{HumanEval} and 90/390 (23\%) for \textsc{MBPP} for \texttt{Gemma-2-2B}, and 40/159 (25\%) instances for \textsc{HumanEval} and 94/390 (24\%) for \textsc{MBPP} for \texttt{Qwen2.5-32B}.

We then applied Group Relative Policy Optimization (GRPO)~\cite{shao2024deepseekmath}, chosen for being on-policy and online, ensuring training samples always reflect the model's current policy.
We used a group size of $G=8$, sampling temperature $\tau=0.5$, and a binary reward function assigning 1 if the solution passes the incremental type checker and 0 otherwise.
We selected this reward function, because it is the simplest reward function that remains agnostic to the specific form of the solution, rewarding only compliance with the constrainer.
To mitigate the risk of catastrophic forgetting, we trained a LoRA adapter~\cite{hu2022lora} ($r=16$, $\alpha=32$) with learning rate $5 \cdot 10^{-5}$ and KL coefficient $\beta=0.04$ until convergence.

After finetuning, we evaluated each model by running it 10 times on the full \textsc{HumanEval} and \textsc{MBPP} datasets at $\tau=0.1$.

\subsubsection{Metrics}

To judge the effect of finetuning, we use the same metrics we used for RQ0, namely, functional correctness, type-syntax correctness, and timeout rate.
We compare how the original models (\textit{OR}) scores against the performance
of the same model finetuned with GRPO (\textit{FT}).
In addition, we also measure the number of generated solutions for each dataset that pass the incremental type checker before and after finetuning.

\subsubsection{Results}
\autoref{tab:results-finetuning} shows the performance of the models before (Columns \textit{OR}) and after (Columns \textit{FT} in gray) finetuning.

In both \textsc{HumanEval} and \textsc{MBPP}, before finetuning \nc is always
significantly better in functional correctness than \constr with a large effect
size (between 3.7 and 10 percentage points across models). 
In \textsc{HumanEval}, after finetuning, \constr and \nc become equivalent for
\texttt{Gemma-2-2B} (around 32\%), with \nc decreasing of 3.7 percentage points;
and become closer for \texttt{Qwen2.5-32B} (around 77\% for \constr and 82\% for \nc),
with \constr increasing of 5.3 percentage points. We attribute the decrease of
\nc to the reward function favouring type-correct solutions that are not
functionally correct (similar to \autoref{lst:forwarded-reference-unconstrained}).
While the increase of \constr can be attributed to an increase alignment,
preventing timeouts (around 3 percentage points less for token limit and 1 to 3
less for time limit across models) and some crashes of the constrainer (around 3
points less for \texttt{Qwen2.5-32B}).
The type-based reward function, however, improves type-syntax correctness of
both strategies for \texttt{Gemma-2-2B} (of around ~8\%), and only of \constr for
\texttt{Qwen2.5-32B} (of around ~5\%).
In particular, the number of solutions generated by \nc that pass the 
incremental type checker increases for all models from 72\% to 92\%
after finetuning, indicating an increased alignment of the models to the
constrainer.

In \textsc{MBPP}, after finetuning, the functional correctness of the \constr and
\nc become closer across models, with an increase of \constr (of around 2 percentage
points for \texttt{Gemma-2-2B} and 6 for \texttt{Qwen2.5-32B}) and a decrease
or very slight increase of \nc (of around 2 percentage points less for 
\texttt{Gemma-2-2B} and 0.6 more for \texttt{Qwen2.5-32B}).
In these cases, however, the functional correctness of \nc remains significantly
better than the one of \constr.
Also similarly to \textsc{HumanEval}, after finetuning, the type-syntax
correctness increases by 8/9\% for \texttt{Gemma-2-2B}, and around 5\% only for
\constr of \texttt{Qwen2.5-32B}.
The number of solutions generated by \nc that pass the incremental type checker increases from 73\% before finetuning to 88\% after. 
Overall, finetuning increases the alignment to the constrainer, but not enough
to make \constr and \nc comparable in terms of functional correctness, contrary
to \textsc{HumanEval}.
An increased alignment is beneficial when it comes to the timeout rate also in \textsc{MBPP},
decreasing max tokens timeouts (of around 2 to 5 percentage points), 
and those due to time limit (of around 2 percentage points), across models.

In both datasets, finetuning increases alignment between the language model and the constrainer and mitigates the incompleteness bias.
Yet, some incompleteness bias remains after finetuning, as evidenced by the persisting timeouts during constrained generation, and a remaining percentage of unconstrained-generated solutions that still do not comply with the constrainer; as a result, constrained decoding continues to underperform the unconstrained strategy, or at best matches it.

\begin{tcolorbox}[boxrule=0pt,frame hidden,sharp corners,enhanced,borderline north={1pt}{0pt}{black},borderline south={1pt}{0pt}{black},boxsep=2pt,left=2pt,right=2pt,top=2.5pt,bottom=2pt]
	\textbf{RQ2 (Restoring Alignment)}: Finetuning increases alignment between the language model and the constrainer across both \textsc{HumanEval} and \textsc{MBPP}, improving type-syntax correctness and reducing timeouts.
	Incompleteness bias persists after finetuning, meaning constrained decoding still underperforms or, at best, matches the unconstrained strategy's functional correctness.
\end{tcolorbox}

\subsection{RQ3: Complete Alignment}

\subsubsection{Procedure}

We evaluated a complete constrainer for code generation and compare its performance to that of several incomplete variants.
As a case study, we considered a syntactic constrainer that is complete with respect to the TOML language specification~\cite{TOML}.
We chose TOML because it is widely adopted, has a simple syntax and semantics, and includes several derived forms~\cite{freund2021union}.
These properties make TOML a good fit for our study: the model is likely to understand it well, and we can easily derive incomplete variants by removing syntactic features or derived forms of the language, without losing expressivity~\cite{schreiter2025novel}. 
The constrainer only enforces syntax: although TOML does not currently include a schema language, the semantic error ``conflicting keys'' is still possible and not captured by the grammar alone (hence $\LT \subseteq \LC$ as in \autoref{fig:venn}.c).

We considered the following constrainer restrictions: 

\begin{itemize}[leftmargin=2em]
	\item \ccomment: forbids the \textit{documentation} feature of comments (e.g., \lstinline[style=tomlstyle]|#comment|).
	\item \cdotted: forbids the \textit{semantic} feature of dotted keys 
	(e.g., \lstinline[style=tomlstyle]|a.b.c = 1|). 
	These are a derived form of inline tables (e.g., \lstinline[style=tomlstyle]|a={b={c=1}}|), thus expressivity is preserved.
	\item \cspaceall: forbids the \textit{syntactic} feature of all optional
	spaces (e.g., \lstinline[style=tomlstyle]|foo=1|).
	\item \cspaceeq: forbids spaces only before the \texttt{=} symbol (e.g., 
	\lstinline[style=tomlstyle]|foo= 1|). 
	This is the smallest constrainer, changing the original grammar by a single lexeme in a production rule.
\end{itemize}

All constrainers are implemented as GGML BNF grammars~\cite{GGMLBNF},
which are used by \texttt{llama.cpp} to perform constrained decoding.
We used \texttt{llama.cpp} to decouple the constraining logic from model deployment.
In particular, we implemented a constraining loop similar to \autoref{alg:constrained-decoding}, and extended \texttt{llama.cpp} with a new API that validates token extensions against the grammar.

Since there are no existing, publicly available, TOML benchmarks, we adapted the \textsc{JsonModeEval} dataset~\cite{JSONModeEval2025}, used for evaluating SynCode~\cite{ugare2025syncode} for TOML generation; we call this new benchmark \textsc{TomlModeEval}.
\textsc{JsonModeEval} contains 100 JSON solutions, each paired with a specification, which includes a JSON schema and natural language instructions.
The task is to generate the JSON solution from the corresponding specification.
In \textsc{TomlModeEval}, we replace the JSON solutions with their equivalent TOML solutions and keep the same specifications, except for minimal adjustments (e.g., asking for a TOML document instead of a JSON object). 

As LLMs, we considered the worst and best language models from RQ0, i.e., \texttt{Gemma-2-2B} and \texttt{Qwen-2.5-32B}, and the more recent \texttt{Qwen-3-Coder-Next-80B} model. 
Each model is executed 10 times on the 100 problems of \textsc{TomlModeEval} with different seeds, temperature 0.1, and a budget of 512 tokens and 60 seconds. 
We then compared statistically (Wilcoxon at $\alpha = 0.05$, Vargha-Delaney effect size $\hat{A}_{12}$~\cite{arcuri2014hitchhikers}) the results of unconstrained decoding with those of constrained decoding, first using the complete grammar and then each of the incomplete variants described above.

\subsubsection{Metrics} 
Type-syntax correctness is measured by parsing the generated TOML documents with the \texttt{tomllib} library in Python's standard library; when the generation times out, either in terms of number of tokens or in terms of time, we consider it a syntax error.
We measured functional correctness in two different ways: exact match (\textbf{EM}) requires the generated document to be identical to the reference solution, and is useful in settings where the generated document is directly consumed by a downstream system; similarity (\textbf{SIM}) is measured as the tree-edit similarity where modifications are weighted by Levenshtein distance, and is useful in settings where the generated document is meant to be consumed by a human (higher similarities indicate that fewer edits are needed).

\subsubsection{Results}

\begin{table}[t]
	\centering
	\caption{\textbf{RQ3.} Results on \textsc{TomlModeEval}, using varying degrees of incomplete constrainers. 
	Bold values indicate statistical significance between \constr and \nc (Wilcoxon at $\alpha = 0.05$) and underlined values a large effect size ($\hat{A}_{12}$).}
	\label{tab:rq3-results-toml}
	\footnotesize
	\setlength{\tabcolsep}{5.2pt}
	\renewcommand{\arraystretch}{1.0}
	\begin{tabular}{llrrrrr}
		\toprule
		\multirow{3}{*}{\textbf{Model}} & \multirow{3}{*}{\textbf{Strategy}} & \multicolumn{2}{c}{\textbf{FC $\uparrow$ (\%)}} & \multicolumn{1}{c}{\multirow{2}{*}{\textbf{TSC $\uparrow$ (\%)}}} & \multicolumn{2}{c}{\textbf{Timeout $\downarrow$ (\%)}} \\
		\cmidrule(r){3-4} \cmidrule(r){6-7}
		&  & \multicolumn{1}{c}{\textbf{EM}} & \multicolumn{1}{c}{\textbf{SIM}} & \multicolumn{1}{c}{} & \multicolumn{1}{c}{\textbf{Tokens}} & \multicolumn{1}{c}{\textbf{Time}} \\

		\midrule

		\multirow{7}{*}{\begin{tabular}[c]{@{}l@{}}\texttt{Gemma-2} \\ \texttt{2B}\end{tabular}} 
		& \nc & 31.3 & 55.2 & 58.9 & 0.0 & 0.0 \\
		& \constr & {\ul \textbf{48.1}} & {\ul \textbf{81.6}} & {\ul \textbf{87.2}} & 8.9 & 0.0 \\
		\cmidrule(r){2-7}
		& \cspaceeq & 28.6 & 36.2 & 37.4 & 62.5 & 0.0 \\
		& \cspaceall & {\ul \textbf{33.9}} & {\ul \textbf{69.2}} & {\ul \textbf{79.0}} & 7.2 & 0.0 \\
		& \ccomment & {\ul \textbf{48.1}} & {\ul \textbf{81.8}} & {\ul \textbf{87.5}} & 9.6 & 0.0 \\
		& \cdotted & {\ul \textbf{48.8}} & {\ul \textbf{81.2}} & {\ul \textbf{86.6}} & 9.6 & 0.0 \\
		
		\midrule

		\multirow{7}{*}{\begin{tabular}[c]{@{}l@{}}\texttt{Qwen-2.5} \\ \texttt{32B}\end{tabular}} 
		& \nc & 40.3 & 77.1 & 89.8 & 0.0 & 0.0\\
		& \constr & {\ul \textbf{41.3}} & {\ul \textbf{80.2}} & {\ul \textbf{93.4}} & 0.0 & 0.0 \\
		\cmidrule(r){2-7}
		& \cspaceeq & 39.7 & {\ul \textbf{82.7}} & {\ul \textbf{92.1}} & 0.0 & 0.0 \\
		& \cspaceall & 29.2 & {\ul \textbf{80.1}} & {\ul \textbf{90.9}} & 0.0 & 0.0 \\
		& \ccomment & {\ul \textbf{41.3}} & {\ul \textbf{80.0}} & {\ul \textbf{93.4}} & 0.5 & 0.0 \\
		& \cdotted & {\ul \textbf{41.3}} & {\ul \textbf{80.0}} & {\ul \textbf{93.2}} & 0.0 & 0.0 \\

		\midrule

		\multirow{7}{*}{\begin{tabular}[c]{@{}l@{}}\texttt{Qwen-3-Coder} \\ \texttt{Next-80B}\end{tabular}} 
		& \nc        & 62.5 & 84.7 & 91.7  & 0.0  & 8.1  \\
		& \constr    & 62.5 & 84.8 & 91.9 & 0.0  & 8.1  \\
		\cmidrule(r){2-7}
		& \cspaceeq  & {\ul \textbf{1.9}}  & {\ul \textbf{2.8}}  & {\ul \textbf{2.9}}  & 10.6 & 86.5 \\
		& \cspaceall & {\ul \textbf{15.7}} & {\ul \textbf{23.1}} & {\ul \textbf{26.2}}  & 2.8  & 71.0 \\
		& \ccomment  & 62.5 & 84.8 & 91.9 & 0.0  & 8.1  \\
		& \cdotted   & 63.0 & 85.4 & 93.6 & 0.0  & 5.7  \\
		\bottomrule
	\end{tabular}
\end{table}

\autoref{tab:rq3-results-toml} shows the results of the sensitivity analysis across all the language models and decoding strategies.
Considering the complete constrainer, i.e., rows \constr in the table, we observe that constrained decoding significantly outperforms the unconstrained strategy for two models, namely \texttt{Gemma-2-2B} and \texttt{Qwen-2.5-32B}, while being equivalent for \texttt{Qwen-3-Coder-Next-80B}.
Constraining (\constr) is more beneficial for smaller models such as \texttt{Gemma-2-2B} (53\% increase w.r.t. \nc for exact match and 48\% for the similarity score), where the unconstrained strategy produces a substantial number of errors (around 40\% with type-syntax errors); interestingly, \texttt{Gemma-2-2B} reaches a higher exact match and similarity score than the much larger \texttt{Qwen-2.5-32B} when both are constrained using the complete syntax.
The benefits gradually decrease as the language model becomes more capable: there is a 2.5\% increase in exact match and a 4\% increase in similarity score for \texttt{Qwen-2.5-32B}, where the unconstrained strategy only produces a syntax error 10\% of the times; \constr and \nc are equivalent for \texttt{Qwen-3-Coder-Next-80B}, where the unconstrained model makes syntax mistakes in only 0.2\% of cases, excluding timeouts.
Also, the complete constrainer does not prevent type errors such as duplicate keys: across 10 runs (1000 generated programs per model), \texttt{Gemma-2-2B} produces 411 type-syntax errors of which 19 are conflicting key errors, \texttt{Qwen-2.5-32B} produces 102 of which 66 are conflicting key errors, and \texttt{Qwen-3-Coder-Next-80B} produces only 2, none of which are conflicting key errors.
Yet, the complete constrainer still improves functional correctness overall.

For \textit{incomplete} constrainers, the most impactful \textit{feature} for language models is the optional space before the equal (\cspaceeq).
For \texttt{Gemma-2-2B}, exact match drops from 48.1\% to 28.6\%, while for \texttt{Qwen-3-Coder-Next-80B} the reduction is far more pronounced, reaching 97\%.
\texttt{Qwen-2.5-32B} appears less sensitive to this feature: exact match remains essentially unchanged at 39.7\% vs. 40.3\% (\nc), though the similarity score still shows a meaningful increase.
Forbidding all optional spaces (\cspaceall) is similarly detrimental, though the reduction is less pronounced than for \cspaceeq alone.
By contrast, removing other features, such as comments (\ccomment) or dotted keys (\cdotted), does not reduce functional correctness relative to the complete constrained strategy (\constr), and this holds consistently across all evaluated models. 
These findings confirm that misalignment is equally detrimental for both type (RQ0) and syntax constraining (RQ3), causing constrained decoding to underperform.

\begin{tcolorbox}[boxrule=0pt,frame hidden,sharp corners,enhanced,borderline north={1pt}{0pt}{black},borderline south={1pt}{0pt}{black},boxsep=2pt,left=2pt,right=2pt,top=2.5pt,bottom=2pt]
	\textbf{RQ3 (Complete Alignment)}:
	Constrained decoding significantly outperforms unconstrained decoding when the model is complete, but incompleteness can reduce functional correctness by up to 97\%. 
	Moreover, optional spaces are more impactful than comments or dotted keys, showing that models attribute semantics to all syntactic constructs, including aesthetic spacing.
\end{tcolorbox}

\section{Threats to Validity} \label{sec:threats}

\paragraph{Internal validity.}
All strategies were evaluated under uniform hardware and timeout conditions, following standard empirical guidelines~\cite{liu2023your}.

\paragraph{Conclusion validity.}
To account for LLM nondeterminism~\cite{chen2021codex, openai2023gpt, ouyang2025empiricalnondeterminism}, we conducted 10 independent runs per configuration. We supported our findings with rigorous statistical tests (Wilcoxon signed-rank and Vargha-Delaney $\hat{A}_{12}$)~\cite{arcuri2014hitchhikers}.

\paragraph{External validity.}
Using a limited number of languages, datasets, and models poses an external validity threat.
In terms of languages, we used TypeScript and TOML, which represent two different constraining models, i.e., types and syntax, respectively.
In terms of datasets, we used \textsc{HumanEval} and \textsc{MBPP} for TypeScript, which are suitable for the Typescript constraining model that we adopted~\cite{muendler2025typeconstrained}, and adapted \textsc{JSONModeEval} for TOML generation, a dataset also used to evaluate existing syntax-constraining approaches~\cite{ugare2025syncode}.
In terms of models, we adopted the six language models from four different families used by prior work~\cite{muendler2025typeconstrained}, and additionally included \texttt{Qwen3-coder-next-80B}~\cite{qwen3_coder_next_tech_report} to cover a more recent and capable model family. 
We evaluated models ranging from 2B to 80B parameters, a range representative of the small-to-medium scale at which constrained decoding is most likely to provide a measurable benefit over unconstrained generation.

\section{Related Work} \label{sec:rel-work}

\paragraph{Constrained Decoding using Grammars} Several works enforce syntactic validity during LLM generation by masking invalid tokens using context-free grammars~\cite{geng-etal-2023-grammar, ugare2025syncode}, optimized DFA masks~\cite{dong2024xgrammar}, or lightweight runtime libraries~\cite{willard2023efficientguidedgenerationlarge, llguidance2025}. 
To address the distribution distortion that is due to masking, \citet{park2024grammaraligned, park2025flexible} propose grammar-aligned decoding to achieve asymptotic probabilistic alignment. 
On the other hand \citet{gonzalez2025mcmc} uses Markov Chain Monte Carlo to iteratively refine valid prefixes that monotonically converge to the unbiased distribution.
Recently, \citet{muendler2025constraineddiffusion} and \citet{suresh2025dingoconstrainedinferencediffusion} proposed grammar constraining for diffusion models that generate tokens in a partially filled sequence in no particular order.
While these studies focus on enforcing context-free syntax rules or aligning probabilities, our work theoretically and empirically investigates when and why constrained decoding is beneficial for code generation.

\paragraph{Complex Constraints} Moving beyond basic syntax, frameworks like Synchromesh~\cite{poesia2022synchromesh} and Monitor-Guided Decoding (MGD)~\cite{agrawal2023mgd} enforce deeper semantic and logical properties using completion engines and static program analysis. 
Similarly, \citet{li2025} integrate solver-guided decoding for formal specifications, and \citet{muendler2025typeconstrained} target type soundness by pruning tokens that violate TypeScript rules. 
\citet{nagy2026chopchop} propose ChopChop, a general framework for constraining the output of an autoregressive language model to follow semantic properties.
Rather than a new constraining approach, our work focuses on modelling the relationships between model language $L_{\mathcal{M}}$, target language $L_T$ and constrained language $L_C$, and on the effects of misalignment between them.

\paragraph{Small Language Models and Constrained Generation} Recent studies emphasize Small Language Models (SLMs) as efficient alternatives to LLMs for code generation, particularly when enhanced by finetuning or structural priors~\cite{belcak2025smalllanguagemodelsfuture, giagnorio2025, hasan2025assessingsmalllanguagemodels}.
Prior literature frames constrained decoding as a helpful crutch to compensate for the reduced capacity of smaller models. 
Our empirical findings reveal that, when the constrainer is misaligned with the target (RQ0---RQ2), SLMs suffer the most from incompleteness bias, hitting timeouts and token limits. 
However, when the constrainer is aligned with the target (RQ3), constrained decoding benefits small language models more (i.e., \texttt{Gemma-2-2B}) than larger ones (i.e., \texttt{Qwen-2.5-32B}).

\section{Conclusion}\label{sec:conclusion}

We show that an incomplete constrainer can be detrimental for code generation.
To address this, we study constrained decoding through the novel lens of \emph{alignment} among the language model, the constraining rules, and the target specification, establishing strict alignment as a fundamental prerequisite for success. 
To improve alignment, we empirically validate two solutions: harmonizing the LLM to the constrainer via finetuning, or aligning the constrainer to the target language, making it complete. 
Our findings uncover the hidden costs of incomplete constraints and provide the community with a novel instrument to advance constrained code generation.

\balance
\bibliographystyle{ACM-Reference-Format}
\bibliography{paper}

\end{document}